\documentclass[prl,reprint, longbibliography,superscriptaddress]{revtex4-1} 

\usepackage{amsmath}  
\usepackage{amsfonts} 
\usepackage{graphicx} 
\usepackage{braket}
\usepackage{textcomp}
\usepackage{gensymb}
\usepackage[T1]{fontenc}
\usepackage{xcolor}

\begin{document}

\title{Spin-Wave Multiplexed Atom-Cavity Electrodynamics}

\author{Kevin C. Cox}
\email[Corresponding author: ]{kevin.c.cox29.civ@mail.mil} 
\affiliation{U.S. Army Research Laboratory, 2800 Powder Mill Rd, Adelphi MD 20783}

\author{David H. Meyer}
\affiliation{U.S. Army Research Laboratory, 2800 Powder Mill Rd, Adelphi MD 20783}
\author{Zachary A. Castillo}
\affiliation{U.S. Army Research Laboratory, 2800 Powder Mill Rd, Adelphi MD 20783}
\affiliation{Department of Physics, University of Maryland, College Park, Maryland 20742, USA}
\author{Fredrik K. Fatemi}
\affiliation{U.S. Army Research Laboratory, 2800 Powder Mill Rd, Adelphi MD 20783}
\author{Paul D. Kunz}
\affiliation{U.S. Army Research Laboratory, 2800 Powder Mill Rd, Adelphi MD 20783}

\date{\today}

\begin{abstract}
We introduce multiplexed atom-cavity quantum electrodynamics with an atomic ensemble coupled to a single optical cavity mode.  Multiple Raman dressing beams establish cavity-coupled spin-wave excitations with distinctive spatial profiles.  Experimentally, we demonstrate the concept by observing spin-wave vacuum Rabi splittings, selective superradiance, and interference in the cavity-mediated interactions of two spin waves.  We highlight that the current experimental configuration allows rapid, interchangeable cavity-coupling to 4 profiles with an overlap parameter of less than 10\%, enough to demonstrate, for example, a quantum repeater network simulation in the cavity. With further improvements to the optical multiplexing setup, we infer the ability to access more than $10^3$ independent spin-wave profiles. 
\end{abstract}

\maketitle

Significant resources are now being devoted to develop intermediate scale quantum systems with tens or hundreds of quantum bits, tunable interactions, and independent control of each element.  Ion traps \cite{debnath_demonstration_2016}, superconducting circuits \cite{ofek_extending_2016}, tweezer arrays of neutral atoms \cite{omran_generation_2019}, and other systems have made exciting recent advances, but scaling precise quantum dynamics from few-body to many-body remains as a primary challenge in quantum science.   

Instead of building up qubit-by-qubit, like the aforementioned platforms, here we focus on a system where quantum information is stored as patterns or images inside a single cavity-coupled atomic ensemble containing up to $10^6$ atoms.  This scalability more closely resembles, for example, that of a neural network, where data is stored and manipulated as patterns and images rather than binary bits \cite{vasilyev_quantum_2008-1, gopalakrishnan_exploring_2012, rotondo_dicke_2015,torggler_quantum_2017}. 

In this Letter, we introduce an apparatus that allows creation of multiple spin-wave excitations with unique spatial profiles.  The spin waves are all collectively enhanced to emit into a single TEM00 cavity mode, and cavity coupling of each spin wave is dynamically controlled using a corresponding Raman dressing beam, generated by a two-dimensional acousto-optic deflector.  Experimentally, we first observe strong spin wave/cavity interactions by measuring a dressed-state vacuum Rabi splitting (VRS) associated with the spin-wave Raman transition.  Second, we discuss how spin waves are protected from cross-talk through collective dephasing, and demonstrate a high degree of distinguishability by observing selective superradiance over the continuum of spin-wave profiles.  Finally, we observe interference as two spin-waves simultaneously interact with the cavity mode.

As a multiplexed atom-cavity interface  \cite{kalachev_multimode_2013}, our system may form the building block of a scalable quantum repeater \cite{kunz_spatial_2019}.  Alternatively, this approach opens an elegant avenue to demonstrate a local bosonic quantum network for efficiently simulating many-body physics \cite{kroeze_spinor_2018, megyeri_directional_2018, clark_interacting_2019}, generating samples from exponentially complex wave functions, or performing entanglement-enhanced or error-corrected quantum sensing.  In the future, our multiplexed atom-cavity system may be combined with nonlinear cavity-mediated interactions or quantum non-demolition measurements that would enable universal quantum operations between spin waves.  The demonstration here, of spin-wave multiplexed cavity electrodynamics, is a significant step toward these goals.

Our apparatus is complementary to an active body of ensemble multiplexing research, using both spatial and spectral degrees of freedom \cite{nunn_multimode_2008, afzelius_multimode_2009-1, lan_multiplexed_2009, golubeva_high-speed_2011, grodecka-grad_high-capacity_2012, sinclair_spectral_2014, pu_experimental_2017, grimau_puigibert_heralded_2017, vaidya_tunable-range_2018, yang_multiplexed_2018, vernaz-gris_highly-efficient_2018}.  Recently, a spatial array of single-photon detectors has been used to resolve 665 independent spin waves in an atomic ensemble \cite{ parniak_wavevector_2017, mazelanik_coherent_2019}.  However, for maximum utility in a quantum network, systems such as this require dynamic routing of free-space output photons. Further, for most previous multiplexing schemes, the interactions (relative to decoherence rates) in free-space ensembles are weak.  Our system presents a path to overcome these two challenges by multiplexing via dynamic classical dressing beams while informational photons emit into a TEM00 optical cavity mode.

\begin{figure*}[t]
\centering
\includegraphics[width=\textwidth]{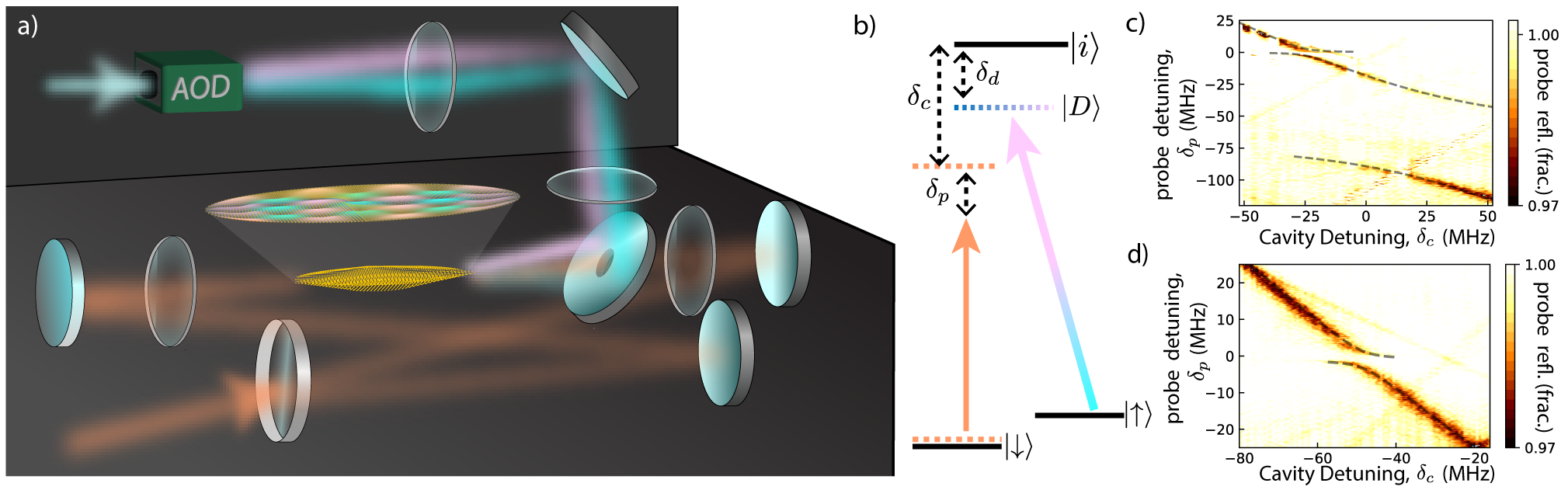}
\caption{Experimental setup.  (a) Rubidium atoms (yellow cloud) are laser-cooled and trapped in an optical ring cavity.  An acousto-optic deflector (AOD) creates one or more dressing laser beams (blue and pink) that impinge upon the cloud at different angles to create spatial spin waves, represented here as a colored grating above the atoms. (b) Level diagram.  The cavity mode (orange) and dressing beam(s) (blue and pink) create two-photon excitations between the long-lived ground states $\ket{\downarrow}$ and $\ket{\uparrow}$.  The dressing laser creates an effective dressed state $\ket{D}$ that interacts with the cavity mode, creating a 2-level system between $\ket{\downarrow}$ and $\ket{D}$. c) A large normal mode splitting is observed in the reflection spectrum (cavity reflection indicated by color), indicating strong collective cooperativity (see text).  We also observe a smaller vacuum Rabi splitting associated with the two-photon transition in part (c) and in the zoomed in dataset of part (d).  For (d), $\delta_d =2\pi\cdot 85$~MHz.  }
\label{figExp}
\end{figure*}

The experimental apparatus is shown in Fig.~\ref{figExp}(a).  We optically trap between $0.3 - 1\times10^6$ rubidium atoms, laser cooled to approximately 20~\micro K, in a 785~nm running-wave optical trap formed by the TEM00 mode of an optical ring cavity (Fig.~\ref{figExp}).  The ring cavity is built in a bowtie configuration using a single 98\% reflective coupling mirror, 3 flat mirrors with reflectivity over 99.8\%, and two anti-reflection coated lenses and vacuum windows.  The ex-vacuo ring cavity follows the design principles outlined in References \onlinecite{cox_increased_2018} and \onlinecite{kunz_spatial_2019}, allowing for large tunability of mode waist and transverse mode splitting while maintaining stability.  The cavity has finesse $\mathcal{F} = 100(10)$, full-width-half-maximum linewidth $\kappa = 2\pi\cdot2.5(1)$~MHz, and free spectral range $FSR = 243(2)$~MHz.  

A level diagram in Fig.~\ref{figExp}(b) shows the dressed-state detuning $\delta_d$, cavity detuning $\delta_c$, and probe-cavity detuning $\delta_p$.  Atoms are initialized in the qubit state $\ket{\downarrow} = \ket{5S_{1/2}, F=1, m_F = 0}$ or $\ket{\uparrow} = \ket{5s_{1/2}, F=2, m_F = 0}$ via optically pumping.  The optical cavity is tuned near resonance between $\ket{\downarrow}$ and the optically excited state $\ket{i}   = \ket{5P_{3/2}, F = 2}$.  Atomic excitations on this transition are characterized by superradiant enhancement to absorb or emit cavity photons, described by the collective VRS, $\Omega_{\mathit{VRS}} = 2 \sqrt{N} g$, where $g$ is the single-atom Jaynes-Cummings coupling parameter and $N$ is the atom number.  This collective coupling is evident by a large normal mode splitting (\emph{i.e.}the VRS) in the cavity reflection spectrum, as shown in Fig.~\ref{figExp}(c).  Figure \ref{figExp}(c) corresponds to $\Omega_{VRS} =2\pi\cdot 67(3)$~MHz, and indicates a large collective cooperativity $N\mathcal{C} = 300(30) \gg 1$, the figure of merit for collective cavity physics.

To create dynamic atom-cavity interactions with the long-lived $\ket{\downarrow} \rightarrow \ket{\uparrow}$ transition, we apply one or more dressing beams at small angles relative to the cavity mode (blue and pink in Fig.~\ref{figExp}).   The dressing beams are generated by a 2-dimensional acousto-optic deflector (AOD) and imaged onto the atomic cloud using two lenses.  

The VRS associated with the dressed state can be seen as a smaller avoided crossing in Fig.~\ref{figExp}(c) for $\delta_d = 2\pi\cdot 65$~MHz.  Figure~\ref{figExp}(d) shows a second spectrogram dataset, zoomed in on the smaller dressed VRS, and taken with a slightly larger dressed-state detuning, $\delta_d = 2\pi\cdot 85$~MHz.  The magnitude of the dressed VRS is given by $\Omega_2^2 = N \mathcal{C} \kappa \gamma_2$ where $\gamma_2$ is the two-photon scattering rate. $\gamma_2 = 2\pi\cdot33(7)$~kHz for the data  in Fig. \ref{figExp}(d). Additional faint structure in Fig \ref{figExp}(c) and (d) is primarily due to higher-order laser sidebands created by the electro-optic modulator used to generate the probing beam.  Being far from atomic resonance, these tones do not interact with the atoms but are nonetheless observed in the cavity spectrum data.  Several discontinuities in the Fig. \ref{figExp}(c) spectrogram are a result of drop-outs in the cavity frequency stabilization electronics as the cavity frequency $\delta_c$ was tuned over a wide range.

A two-photon excitation (from $\ket{\downarrow}$ to $\ket{\uparrow}$) generated by a dressing field and a cavity-mode photon generates coherence in the atomic state described by a collective spin operator (\emph{i.e.} Bloch vector) $\vec{\textbf{J}}^{k}$ with an arbitrary holographic spatial profile.  This profile is defined by the excitation's wave vector $\vec{k}(\vec{r})$ where $\vec{r}$ is the position vector in the lab frame ($\vec{k}$ vector overhat symbols omitted in subscripts, superscripts, and to denote magnitude).  The collective Bloch vector can be written in terms of spatially dependent raising and lowering operators and the collective $\textbf{J}_z$ operator. $\textbf{J}^{k}_\pm =  \frac{1}{2}\sum_i^N e^{\pm i \phi^k_i} \sigma^i_\pm$, and $\textbf{J}^{k}_z = \frac{1}{2}\sum_i^N \sigma^i_z$ where $\sigma^i_\pm$ and $\sigma^i_z$ are the standard Pauli spin operators for the $i$th atom.  The phase pattern of a given spin wave, in general, can be any holographic profile, $\phi^{k}_i = \vec{k}(\vec{r}_i) \cdot \vec{r}_i$. This spatial dependence is the key element that allows multiplexing of collective atomic excitations in this system.  

\begin{figure}[t]
\centering
\includegraphics[width=\columnwidth]{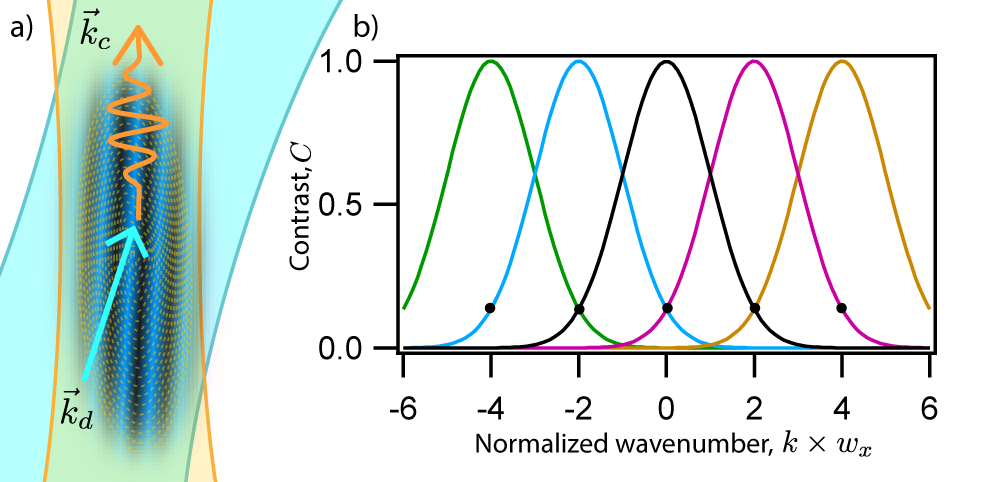}
\caption{Spin-wave multiplexing  (a) A two-photon excitation creates a spatial phase grating (blue and black stripes).  (b) Multiplexing occurs in momentum space parameterized by wave vector $\vec{k} = \vec{k}_d-\vec{k}_c$.  Spin waves created at specific $k\times w$ (-4 to 4 shown), are only collectively enhanced to interact with the cavity within a Gaussian width set by $w_x$, the transverse width of the atom cloud.  The overlap parameter between two neighboring spin waves is represented by the points labeled with black dots.}
\label{figSW}
\end{figure}

Destructive interference causes collective excitations created with a specific initial spatial grating labeled $\vec{k}_1$ to be protected from 2-photon cavity-interactions based on another arbitrary phase grating $\vec{k}$.  One may define a unitary operator that transforms between the $\vec{\textbf{J}}^{k_1}$ and $\vec{\textbf{J}}^k$ Bloch vectors, $\textbf{J}^{k}_\pm = \textbf{U}^{k}_{k_1} \textbf{J}^{k_1}_\pm$,  where the transformation re-orients the phase of each atom's spin operators, $\textbf{U}^{k}_{k_1} = \sum_{i=1} ^N \mathbf{R}_{zi}(\phi^{{k_1}}_i-\phi^{k}_i).$  $\mathbf{R}_{zi}(\alpha)$ is a rotation operator for the $i$th atom's Bloch vector about the $z$ spin axis by amount $\alpha$.  The result is that the superradiant collective spin of the $\vec{k}_1$ exciation becomes non-superradiant in the new $\vec{k}$ profile.  Building on previous work \cite{cox_reducing_2014}, the residual transverse coherence (that leads to the superradiant enhancement), as perceived in an arbitrary spatial profile $\vec{k}$, is $C_d(\vec{k}) \equiv  J_\perp^{k}/ J_\perp^{k_1}$, where $\perp$ denotes the magnitude of transverse projection of the Bloch vector \cite{cox_reducing_2014}.  This reduction in coherence serves as the numerical factor that describes the low sensitivity of the $\vec{k}_1$ excitation to coherent operations in the $\vec{k}$ spatial profile. 

In our experimental realization of angular multiplexing, the cavity mode and dressing beam are approximately plane waves, applied at a small relative angle $\theta$ (\emph{i.e.} $k= |\vec{k_d} - \vec{k_c}| \approx \theta \times k_0 $ where $k_0 = 2 \pi/\lambda$ and $\lambda = 780~$nm here.).  When $\theta$ is changed by a small amount $\Delta \theta$ from the original value $\theta$, the transverse coherence is expected to be $C_d(\Delta \theta) = e^{-\Delta \theta^2 k_0^2 w_x^2 /2}$ for our atomic ensemble, with a Gaussian density profile of width $w_x$. 

In Fig.~\ref{figSW}(b), we plot $C_d$ for five hypothetical spin waves with $ k \times w$ ranging between -4 and 4.  This plot graphically represents the multiplexing in $k$ space, with an overlap factor $O$ between two spin-wave profiles $\vec{k}_1$ (\emph{i.e.} $\theta_1$) and $\vec{k}_2$ ($\theta_2$) defined as the coherence value at the location of the neighboring spin-wave $O \equiv C_d(\theta_2 - \theta_1)$. For the case shown, $O = 0.135$  (points denoting the overlap factor between neighboring spin waves are plotted as black dots in the figure).  

The maximal optical multiplexing of the system is limited by the optical wavelength, $k_{\text{max}} \approx 2 k_0 = 1.6\times10^7$ $\text{m}^{-1}$.  The quantum limit to the multiplexing, for unentangled atoms, is derived from the standard quantum limit for coherent spin states \cite{wineland_squeezed_1994} and nominally leads to a minimum overlap factor $O_{min} = 1/\sqrt{N}$. These limitations to $k$ and $O$ can be surpassed using atom-atom entanglement and optical superresolution, respectively.

We experimentally measure the degree of overlap between two spin-wave profiles as a function of dressing angle by observing selective superradiance \cite{asenjo-garcia_exponential_2017}.  The atoms are optically pumped into the $\ket{\uparrow}$ state, and a dressing laser is applied with the cavity on two-photon resonance ($\delta_p = 0$).  The atoms collectively emit a strong superradiant light pulse into the cavity mode.  The superradiant pulse is nominally seeded by vacuum fluctuations that lead to a random global emission phase as well as fluctuations in the timing of the pulse.  Most importantly, the superradiance is characterized by the spontaneous buildup of large collective coherence into the spin-wave $\vec{\textbf{J}}^{k_1}$ set by the applied dressing laser.  It is worth noting that superradiant emission in similar systems has recently been shown to serve as the basis for ultra-narrow linewidth lasers \cite{norcia_frequency_2018}.

We turn on the initial dressing laser (defining spin wave $\vec{k}_1$) at time $t=0$ and allow collective coherence to be spontaneously established through the buildup of a superradiant pulse.   At time $t=3$~\micro s, we abruptly (at the speed of the AOD, approximately $0.6$~\micro s) change the angle of incidence $\theta$ of the dressing beam.  For this test, we apply equivalent frequency changes of opposing sign to the two dimensions of the AOD, so the overall dressing laser frequency does not change.  This operation selects a second phase profile $\vec{k_2}$.  The overlap parameter $O$ between the $\vec{k}_1$ and ${k}_2$ profiles quantifies the amount of superradiant light that will be emitted directly after the change, since instantaneous emitted power of a superradiant ensemble is $P \sim \mathcal{C} \gamma_2 J_\perp^2$ \cite{bohnet_linear-response_2014}.  This means that the power emitted after the angle change $\Delta \theta$ is expected to be $P  \sim \mathcal{C} \gamma_2 N^2 O^2$.

\begin{figure}[t]
\centering
\includegraphics[width=\columnwidth]{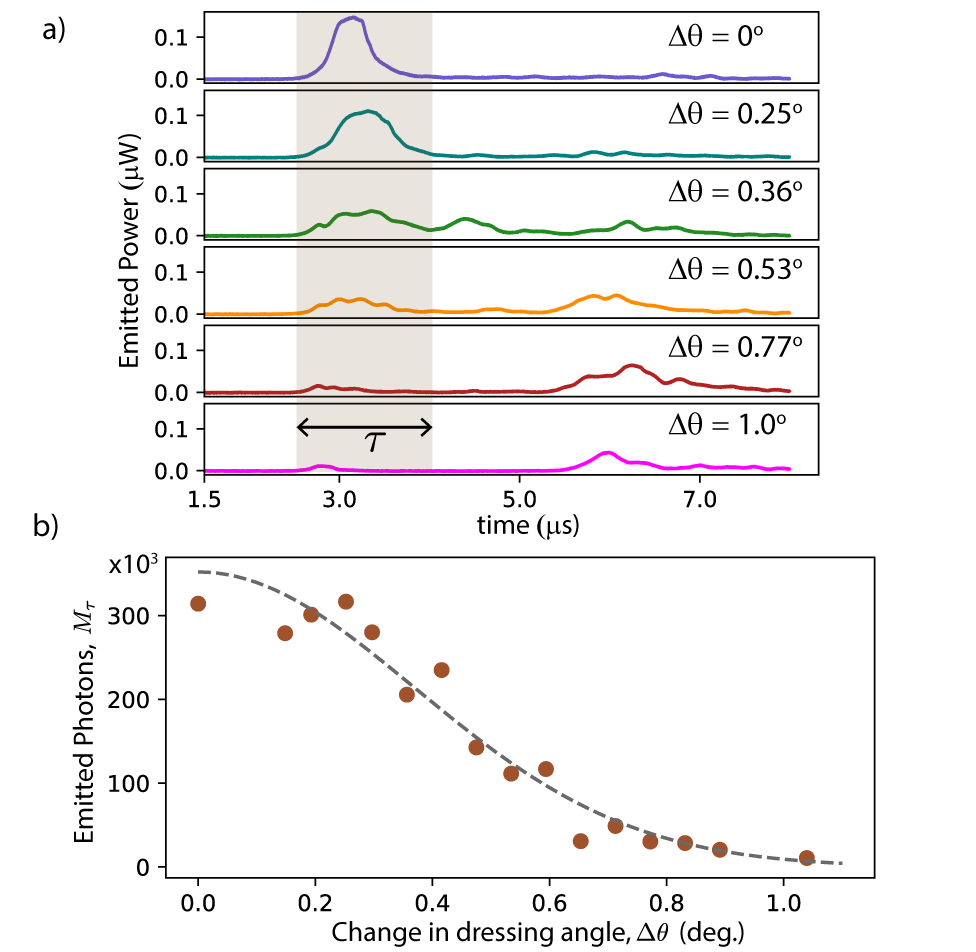}
\caption{Selectve superradiance. (a) Superradiant emission from atoms in the $\ket{\uparrow}$ state, characterized by strong emission pulses at approximately $t= 3~$\micro s.  At $t=3$~\micro s, we abruptly change the angle of the dressing laser using the AOD.  The total detected power in the shaded region ($M_\tau$) may be related to the overlap $O$ between the two spin wave profiles. (b)  $M_\tau$ plotted versus the angle change $\Delta \theta$ of the dressing beam.  The Gaussian decay quantifies the overlap parameter and therefore the spin-wave capacity of the ensemble.}
\label{figSR}
\end{figure}

The data for this experiment are shown in Fig.~\ref{figSR}(a).  The averaged emitted power from the cavity is plotted versus time for six angle changes $\Delta \theta$.  The average delay between turn-on and the pulse emission is approximately $t_\text{delay} \approx 3$~\micro s, governed by the rate, $N \mathcal{C} \gamma_2$, of non-collective emission into the cavity.  We post-select trials where an initial superradiant pulse begins at approximately $t = 2.8$~\micro s, and average over approximately 10 post-selected trials for each angle setting. We define a time window $\tau$ (shaded in \ref{figSR}(a)) in which the emitted power quantifies the squared state overlap $O^2$.  For small angle changes $\Delta \theta$, we predict $M_{\tau} \propto O^2 \sim e^{-\Delta \theta^2  k_0^2 w^2}$, where $M_{\tau}$ is the number of 
detected photons in the time window $\tau$.

In Fig. \ref{figSR}(a) $M_{\tau}$ decreases as $\Delta \theta$ increases.  But interestingly, even for the cases where the overlap is greatly reduced, the atomic excitation eventually leaves the cavity by re-establishing superradiance in the new $\vec{k}_2$ spatial profile.  This is clear in the data of Fig.~\ref{figSR}(a) as a second pulse is established, on average, at time $2 t_\text{delay} \approx 6$~\micro s.  The total emitted energy begins to decrease at and above $\Delta \theta = 1^\circ$, most likely due to alignment errors and limited range in the current optical multiplexing apparatus.

In Fig. \ref{figSR}(b) we plot $M_{\tau}$ as a function of $\Delta \theta$.   The fitted decay of the initial superradiant pulse energy follows a Gaussian decay as expected, corresponding to a transverse Gaussian width of the ensemble of $w = 14(1)$~\micro m, in reasonable agreement with the expectation based on atom temperature and trap depth.

\begin{figure}[t]
\centering
\includegraphics[width=\columnwidth]{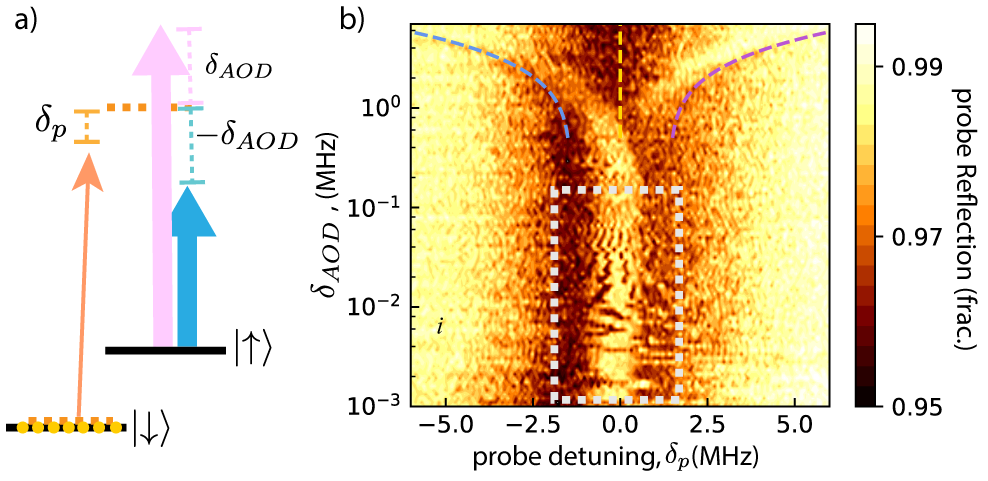}
\caption{Interference between two simultaneous spin waves.  (a) Level diagram.  Two dressing lasers are applied at different detunings $+\delta_{AOD}$ and $-\delta_{AOD}$, and angles $\theta = 0.21 \times \delta_{AOD}$~rad.  (b) Cavity reflection spectrogram.  The system's calculated normal modes in steady state are plotted as blue, orange, and pink dashes.  At smaller values of $\delta_{AOD}$, we observe beating in the atom-cavity interaction (seen as stripes near $\delta_a = 0$, inside the white box) as the two spin-waves' cavity interactions constructively and destructively interference.  }
\label{figInt}
\end{figure}

As a further investigation of spin-wave interference, we observe the cavity dynamics of two spin waves interacting with the cavity mode at the same time.  A level diagram for this experiment is shown in Fig.~\ref{figInt}(a).  We initialize all atoms into the $\ket{\downarrow}$ state, and probe the cavity reflection spectrum (Fig. \ref{figInt}(b)) with $\delta_c = 0$.  We simultaneously excite two spin wave profiles at different angles of incidence and different frequencies ($\pm \delta_{AOD}$), by applying appropriate tones to the AOD.

In each experimental run, the probe detuning $\delta_p$ is scanned from -5~MHz to 5~MHz over 0.4~ms.  At small $\delta_{AOD}$, we observe a dressed VRS similar to Fig.~\ref{figExp}(d).  However, the normal-mode splitting oscillates due to constructive and destructive interference of the spin waves as their relative phases beat against each other in time.  The interference is visible inside the white dashed box of Fig.~\ref{figInt}(b) as a chirped stripe pattern. This qualitatively verifies the coherence of the cavity-mediated exchange interaction between the two simultaneous spin waves, and may form the basis for future experiments to create a wide array of entangled states between multiple spin waves.

When the $\delta_{AOD}$ becomes large, along with the beam's relative angle of incidence, the individual spin-wave excitations split apart from the cavity-like mode in the spectrum, and become distinct.  In the large $\delta_{AOD}$ regime, dashed lines display the theoretical normal mode structure calculated by diagonalizing the Jaynes-Cummings Hamiltonian.  These data represent the two spin-wave excitations becoming distinct in both spectral frequency and spatial overlap.

The current configuration of the experiment uses small incidence angles that allow for longer spin-wave wavelengths, reducing the sensitivity to thermal motion of the atoms with respect to the spin grating, and thus increasing the spin-wave lifetime.  In this configuration we infer, from the data of Fig. \ref{figSR} and the available degrees of freedom, the ability to rapidly access four spin-waves with an overlap parameter of less than 10\%. In the future, we may instead address the atoms perpendicular to the cavity axis and thereby provide access to more than $10^3$ independent spin waves with expected overlap of less than 1\%.   To improve spin-wave lifetime in the high-capacity scenario, an optical lattice may be added. However, the current system with multiple independent spin waves may already allow for a proof-of-principle simulation/demonstration of a quantum repeater within a distance-scalable quantum network. Future experiments will explicitly evaluate cross-talk in multiple spin wave dynamics and longer coherence times.  Pushing the capability of this platform further will involve implementing a universal set of entanglement-generating cavity interactions using cavity feedback \cite{leroux_implementation_2010}, quantum non-demolition measurements, or Rydberg states.  Armed with these tools and concepts, we hold optimism that multiplexed cavity quantum electrodynamics will be a promising quantum many-body platform for networking, entanglement-enhanced quantum sensing, or even quantum computing at the intermediate scale.

\begin{acknowledgments}
The authors would like to thank Michael Foss-Feig, Siddhartha Santra, James Thompson, Alexey Gorshkov, Kanupriya Sinha,  Elizabeth Goldschmidt, Joseph Britton, and Paul Lett for useful discussions and advice.   

\end{acknowledgments}

\bibliography{main}

\end{document}